\documentclass[preprint,aps,prb]{revtex4}
\usepackage[dvips]{graphicx}
\usepackage{color,array,dcolumn}
\usepackage{amsmath}
\usepackage{amssymb}

\begin{document}

\title{Van der Waals Interactions in Density Functional Theory by combining the 
Quantum Harmonic Oscillator-model with Localized Wannier Functions} 

\author{Pier Luigi Silvestrelli}
\affiliation{Dipartimento di Fisica e Astronomia, 
Universit\`a di Padova, via Marzolo 8, I--35131, Padova, Italy,
and DEMOCRITOS National Simulation Center, of the Italian Istituto 
Officina dei Materiali (IOM) of the Italian National 
Research Council (CNR), Trieste, Italy}


\begin{abstract}
\date{\today}
We present a new scheme to include the van der
Waals (vdW) interactions in approximated Density Functional Theory (DFT)
by combining the Quantum Harmonic Oscillator model with 
the Maximally Localized Wannier Function technique.
With respect to the recently developed DFT/vdW-WF2 method,
also based on Wannier Functions, the new approach is more
general, being no longer restricted to
the case of well separated interacting fragments.
Moreover, it includes higher than pairwise energy contributions,
coming from the dipole--dipole coupling among quantum oscillators.
The method is successfully applied to the popular S22 molecular database, and
also to extended systems, namely graphite and H$_2$ adsorbed on the
Cu(111) metal surface (in this case metal screening effects are taken into
account).
The results are also compared with those obtained by 
other vdW-corrected DFT schemes.
\end{abstract}

\maketitle

\section{Introduction}
Density Functional Theory (DFT) is a well-established
computational approach to study
the structural and electronic properties of
condensed matter systems from first principles.
Although current, approximated density functionals 
allow a quantitative description
at much lower computational cost than other 
first principles methods, they fail\cite{Kohn} to properly describe
dispersion interactions. Dispersion forces originate from 
correlated charge oscillations in
separate fragments of matter and the most important 
component is represented by the $R^{-6}$ van der Waals (vdW) 
interaction,\cite{london} originating from correlated instantaneous dipole
fluctuations.
This kind of interactions play a fundamental role in determining the
structure, stability, and function of a wide variety of systems,
including molecules, clusters, proteins, nanostructered materials,
molecular solids and liquids, and in adsorption processes of fragments weakly
interacting with a substrate ("physisorbed").

In the last few years a variety of practical methods have been proposed
to make DFT calculations able to accurately describe vdW effects (for a
recent review, see, for instance, refs. \onlinecite{Riley,MRS,Klimes}).
In these respect, a family of such methods, all based
on the Maximally Localized Wannier Functions (MLWFs),\cite{Marzari}
has been developed, namely the original 
DFT/vdW-WF,\cite{silvprl,silvmetodo,Mostofi}
DFT/vdW-WF2\cite{C3} (based on the London expression and taking into account
the intrafragment overlap of the MLWFs), and
DFT/vdW-WF2s\cite{PRB2013} (including metal-screening corrections), 
successfully applied to a variety of systems:
\cite{silvprl,silvmetodo,silvsurf,CPL,silvinter,ambrosetti,C3,Ar-Pb,Costanzo,Ambrosetti2013,PRB2013,Mostofi}
small molecules, water clusters, 
graphite and graphene, water layers interacting with graphite, 
interfacial water on semiconducting substrates,
hydrogenated carbon nanotubes, 
molecular solids, the interaction of rare gases and small molecules 
with metal surfaces,...
 
All these methods rely on the possibility of attributing the 
MLWFs to well separated fragments.
Although this requirement
is certainly satisfied in many interesting cases (interaction 
between saturated molecules, adsorption of rare gases on metal 
substrates,...), nonetheless it represents both a fundamental
limitation and a practical technical complication, since 
the different fragments must be somehow identified.

In the present paper we overcome the above limitation by presenting 
a method, also based on the MLWFs, developed by adopting the
coupled Quantum Harmonic Oscillator 
(QHO) model.\cite{Cao,Donchev,Tkatchenko12,Reilly,PNAS,QHO} 
This also provides an effective many body description of the 
long-range correlation energy,
beyond a pairwise $C_6/R^6$ approximation.
The new method, hereafter referred to as DFT/vdW-QHO-WF, is successfully 
applied to the popular S22 benchmark set\cite{Jurecka} of weakly interacting
molecules and also to extended systems, namely graphite 
and H$_2$ adsorbed on the
Cu(111) metal surface (in this case metal screening effects are taken into
account).
The results are compared with those obtained by 
other vdW-corrected DFT schemes and with available, reference experimental
values.

\section{Method}
For a system of $N$ three-dimensional QHOs the exact total energy can be 
obtained\cite{Cao,Donchev,Tkatchenko12,Reilly,PNAS,QHO} 
by diagonalizing the $3N \times 3N$ 
matrix $C^{QHO}$, containing $N^2$ blocks $C_{ij}^{QHO}$ of size $3 \times 3$:

\begin{equation}
C_{ii}^{QHO} = \omega_i^2{\bf I}\,\,\,\, ; \,\,\,\,
C_{i\neq j}^{QHO} = \omega_i\omega_j{\sqrt {\alpha_i\alpha_j}}T_{ij}
\label{CQHO}
\end{equation}

where ${\bf I}$ is the unitary matrix, and
$\omega_i$ and $\alpha_i$ are the characteristic frequency and
the static dipole polarizability, respectively, of the $i$-th oscillator.
The interaction (correlation) energy is given by the 
difference between the square root of the eigenvalues
{$\lambda_p$} of the {\it coupled} system of QHOs and of the eigenvalues of the
{\it uncoupled} system of QHOs (namely the characteristic frequencies):

\begin{equation}
E_{c,QHO} = 1/2 \sum_{p=1}^{3N} \sqrt{\lambda_p} - 3/2 \sum_{i=1}^{N}\omega_i
\,.
\label{interac}
\end{equation}

The so-computed interaction energy naturally includes many body
energy contributions, due to the dipole--dipole coupling among the QHOs.
Moreover, it can be proved \cite{QHO} that, within the present model,
the ACFDT-RPA (adiabatic connection fluctuation 
dissipation theorem-random phase approximation) 
correlation energy coincides with the exact binding energy $E_{c,QHO}$.
Hence, the QHO model provides an efficient description of the
correlation energy for a set of localized fluctuating dipoles at an effective
RPA level.

In this paper, we combine the QHO model with the MLWF technique by assuming that each
MLWF can be represented by a three-dimensional harmonic oscillator, so that
the systems is described as an assembly of fluctuating dipoles.
Following ref. \onlinecite{QHO} the bare Coulomb potential is modified
to account for orbital overlap at short distances (thus introducing
a short-range damping):

\begin{equation}
 V_{ij}=\frac{{\it erf}(r_{ij}/\sigma_{ij})} {r_{ij}}\;,
\label{Vij}
\end{equation}

where $r_{ij}$ is the distance between the $i$-th and the $j$-th 
Wannier Function Center (WFC), and  $\sigma_{ij}$ is an effective width,
$\sigma_{ij}=\sqrt{S_i^2 + S_j^2}$, where $S_i$ is the spread 
of the $i$-th MLWF.  
Then, in Eq. \eqref{CQHO} the dipole interaction tensor is\cite{QHO}

\begin{equation}
T_{ij}^{ab} = -\frac{3r_{ij}^ar_{ij}^b-r_{ij}^2\delta_{ab}} {r_{ij}^5}
\left( {\it erf}(\frac {r_{ij}} {\sigma_{ij}})-
\frac{2} {\sqrt{\pi}}\frac{r_{ij}} {\sigma_{ij}}
e^{-(\frac{r_{ij}} {\sigma_{ij}})^2} \right) +
\frac{4} {\sqrt{\pi}}\frac{1} {\sigma_{ij}^3}
\frac{r_{ij}^ar_{ij}^b} {r_{ij}^2}
e^{-(\frac{r_{ij}} {\sigma_{ij}})^2} 
\label{T}
\end{equation}

where $a$ and $b$ specify Cartesian coordinates ($x,y,z$), 
$r_{ij}^a$ and $r_{ij}^b$ are the respective components of the
distance $r_{ij}$, and $\delta_{ab}$ is the
Kronecker delta function.

Moreover, as in ref. \onlinecite{C3}, adopting a simple classical theory,
the polarizability
of an electronic shell of charge $eZ_i$ and mass $mZ_i$, tied to a heavy
undeformable ion is written as
                                                                                
\begin{equation}
\alpha_i = \zeta \frac{Z_i e^2}{m\omega_i^2}\,.
\label{alfa}
\end{equation}

Then, given the direct relation between polarizability and 
volume,\cite{polvol} we assume that $\alpha_i\sim \gamma S_i^3$,
where $\gamma$ is a proportionality constant, so that the orbital volume is
expressed in terms of the $i$-th MLWF spread, $S_i$.

Similarly to ref. \onlinecite{QHO}, we combine 
the QHO model, which accurately describes the long-range
correlation energy, with a given semilocal, Generalized Gradient Approximation
(GGA), functional (PBE\cite{PBE} in
our case), which is expected to well reproduce short-range correlation
effects, by introducing an empirical parameter $\beta$ that
multiplies the QHO-QHO interaction parameter $\sigma_{ij}$ in Eq. \eqref{Vij}.
The three parameters $\beta$, $\gamma$, and $\zeta$ are set up by minimizing the
mean absolute relative errors (MARE), measured with respect
to high-level, quantum-chemistry reference values relative to the
S22 database (see below). 
By taking PBE\cite{PBE} as the reference DFT functional, we get: 
$\beta=1.39$, $\gamma=0.88$, and $\zeta=1.30$.   
In our previous DFT/vdW-WF2\cite{C3} approach, 
the constant $\gamma=0.87$ was instead set up by imposing that 
the exact value for the H atom polarizability
($\alpha_H=$4.5 a.u.) is obtained (of course, in the H case, one
knows the exact analytical spread, $S_i=S_H=\sqrt{3}$ a.u.).
It is clearly reassuring that,
in spite of the different adopted recipes, the resulting
$\gamma$ values are very similar. 
  
Note that, differently from ref. \onlinecite{QHO}, here the atom-based 
point of view is replaced by an electron-based point of view, 
so that the approach is also
applicable to systems, such as metals and semimetals, which cannot
be described in terms of assemblies of atoms only weakly perturbed
with respect to their isolated configuration. 
Moreover, differently from the previous DFT/vdW-WF and DFT/vdW-WF2 schemes,
in the new DFT/vdW-QHO-WF method it is no longer necessary to subdivide the
system into separated fragments, and the certain degree of empiricism 
associated to the presence of a suitable damping function is now eliminated. 

In the specific case of adsorption on metal surfaces, a proper treatment
of metal screening effects is 
mandatory.\cite{Ruiz,Tkatchenko12,Reilly,Klimes,Cole,PRB2013}
Although the QHO model intrinsically describes many--body effects in
the case of localized fluctuating dipoles, the present approach is not 
well suited for describing the response of delocalized
electrons. 
In fact, for MLWFs characterized by large spread, the single QHO approximation
is less appropriate, as the response of delocalized electrons is
expected to be closer to that of a homogeneous electron gas.\cite{Maggs}
To this aim, for the H$_2$-Cu(111) system, we adopted two 
different recipes: a ``single-layer''
approximation, similar to the simplest scheme
(DFT/vdW-WF2s3) among those proposed in ref. \onlinecite{PRB2013},
and a Thomas-Fermi scheme.

In the ``single-layer'' approach,
essentially, vdW effects are only restricted to the interactions of the
adparticle with the topmost metal layer;\cite{Hanke} in fact,
as a consequence of screening, one expects that the 
topmost metal atoms give the dominant contribution.   
We have implemented this by introducing fractional occupation numbers
assigned to the MLWFs, as suitable weights to describe screening
effects in the metal substrate. In particular, the occupation number
assigned to the $i$-th WFC is given by:

\begin{equation}
f_{i} = 2\left(1-{\frac {1}{1 + e^{(z_i-z_r)/{\Delta z}}} }\right) \;,
\end{equation}

where $z_i$ is the vertical coordinate of the $i$-th WFC, 
the reference level $z_r$ is
taken as the level of the second, topmost surface layer, and we assume that
$\Delta z = $(interlayer separation)$/4$; we found that the estimated
equilibrium binding energies and adparticle-surface distances exhibit
only a mild dependence on the $\Delta z$ parameter.

Instead in the Thomas-Fermi scheme the potential of Eq. \eqref{Vij} is
replaced by

\begin{equation}
 V_{ij}=\frac{{\it erf}(r_{ij}/\sigma_{ij})\, e^{-q r_{ij}} } {r_{ij}}\;,
\label{VijTF}
\end{equation}

where $q$ is the standard Thomas-Fermi wave vector, $k_{TF}$, appropriate for
the Cu bulk metal if both 
the $i$-th and the $j$-th WFC are inside the metal slab, $q=0$
if both the WFCs are outside the metal slab, while, in the 
intermediate cases, $q=k_{TF}\, r_{ij}^{in}/r_{ij}$, that is 
$k_{TF}$ is renormalized by considering the portion, $r_{ij}^{in}$, of the
$r_{ij}$ segment which is inside the metal slab. 

Therefore, the present DFT/vdW-QHO-WF method includes both a 
short-range damping (to take orbital overlap effects into account) and,
where appropriate, a long-range damping (to take metal-screening effects 
into account).

The calculations have been performed 
with both the CPMD\cite{CPMD} and the Quantum-ESPRESSO ab initio 
package\cite{ESPRESSO}
(in the latter case the MLWFs have been generated as a post-processing 
calculation using the WanT package\cite{WanT}). 
Electron-ion interactions were described using norm-conserving
pseudopotentials and the PBE reference DFT functional\cite{PBE}
which was adopted also in ref. \onlinecite{QHO} and represents 
one of the most popular GGA choices.

\section{Results and Discussion}
In order to assess the accuracy of the DFT/vdW-QHO-WF method
we have chosen to start by considering the
S22 database of intermolecular interactions,\cite{Jurecka} a widely
used benchmark database, consisting of weakly interacting molecules
(a set of 22 weakly interacting dimers mostly of biological importance), 
with reference binding energies calculated
by a number of different groups using high-level
quantum chemical methods. In particular, we
use the recent basis-set extrapolated CCSD(T) binding
energies calculated by Takatani et al.\cite{Takatani} These binding
energies are presumed to have an accuracy of about 0.1
kcal/mol (1\% relative error). 
Table I summarizes
the results of our calculations on the S22 database, compared to those
obtained by other vdW-corrected DFT schemes, namely 
our previous DFT/vdW-WF\cite{silvprl,silvmetodo,Mostofi} 
and DFT/vdW-WF2\cite{C3} methods,
vdW-DF,\cite{Dion,Langreth07} vdW-DF2,\cite{Lee-bis} VV10\cite{Vydrov} and
rVV10\cite{Sabatini}
(the revised, computationally much more efficient version
of the VV10 method), PBE+TS-vdW,\cite{TS} and PBE+MBD.\cite{Tkatchenko12}
For the sake of completeness we also report data relative to the semiempirical
PBE-D3\cite{Grimme} approach and to the bare, non-vdW-corrected, PBE 
functional.
As can be seen, considering the MARE, the performances of the 
DFT/vdW-QHO-WF method are good, being only
inferior to those of the rVV10, VV10, and PBE+MBD schemes, 
which nowadays probably
represent the most accurate vdW-corrected DFT approaches for 
noncovalently bound complexes.\cite{Sabatini,Tkatchenko12}  
In particular, the mean absolute error (MAE) of DFT/vdW-QHO-WF
(31 meV) is well below the so-called "chemical accuracy" threshold
of 1 kcal/mol (about 43 meV), required to attribute a genuine
quantitative character to the predictions of an ab initio scheme. 

These findings are certainly very promising, however, since the
three parameters $\beta$, $\gamma$, and $\zeta$ have been set up by 
just minimizing the MARE relative to the S22 database (see above),
in order to give a stronger assessment of the quality of the method,
further tests are required, to check whether good performances are
obtained also for other interesting systems, always keeping the
same values of the parameters.  
Therefore we have considered graphite, whose interlayer separation 
distance and binding energy are notoriously difficult quantities to
reproduce by standard DFT approaches. 
In our calculations graphite was modeled with two graphene layers,
with A--B stacking (so, to be precise, the calculations refer
to two isolated graphene layers and not to a bulk graphite solid);
the periodic cell contained 144 C atoms, and an empty region 
of about 16 \AA\ width was left among the graphite replicas, 
in the direction orthogonal to the graphene planes. The 
in-plane geometry was fixed to the one determined experimentally
(C--C distance = 1.421 \AA), while the vertical, interlayer distance
was optimized. The sampling of the Brillouin Zone was limited to
the $\Gamma$ point and, for the DFT/vdW-QHO-WF method, the reference
DFT functional was again PBE. 
Since graphite is a periodically repeated system, each C atom
is subject to vdW interactions due not only to the other C atoms
contained in the reference supercell, but also to their periodic
images. In order to take this effect into account and, at the same time,
avoid to have to diagonalize too large QHO matrices, the DFT/vdW-QHO-WF 
method was applied only to interactions within the reference supercell,
while accounting for the interaction with the periodic replicas.

We define the binding energy (per C atom) as 
$E_b = (E_2-2E_1)/N_C$, where $E_2$ and $E_1$ are the total energies
relative to the two graphene layers (at the optimized
equilibrium distance) and to a single graphene layer,
respectively, and $N_C=144$ is the total number of C atoms contained in 
the simulation supercell.
Note that, in the literature, alternative definitions for the interlayer
binding energy of graphite exist,\cite{Kack,Spanu} also considering the
relationship with the actual experimental measurements:
the ``exfoliation'' energy, $E_e$, is the energy required to
remove one graphene plane from the surface of a graphite solid, while
the ``cleavage'' energy, $E_c$, refers to the interaction between 
two semi-infinite graphite crystals. 
In any case, the knowledge of $E_b$ at the equilibrium interlayer distance
and of the binding energy at a second-layer distance, $E_{bs}$, allows
to estimate\cite{Kack} $E_e$ and $E_c$: 
$E_e = E_b + E_{bs}$, $E_c = E_b + 2E_{bs}$.  

No real direct measurement of the interlayer binding energy of
graphite has been performed so far and the experimental estimates 
are quite scattered, ranging\cite{Spanu,Benedict,Zacharia,Liu} 
from about -60 to -20 meV/atom.
At present, probably the most reliable experimental reference value
has been obtained from desorption experiments on polyaromatic molecules
from a graphite surface,\cite{Zacharia} leading to an estimated value of the
interlayer binding energy and of the cleavage energy of -52 $\pm 5$ and
of -61 $\pm 5$ meV/atom, respectively. 

In Table II the interlayer binding energies and equilibrium distance 
of graphite, computed by DFT/vdW-QHO-WF, are compared to data obtained
by other theoretical schemes, including a Quantum Monte Carlo 
approach,\cite{Spanu} which is a many-body technique able to account
for vdW interactions, and also to some experimental estimates 
(theoretical data are not corrected by zero-point motion
and lattice vibrational contributions).
As can be seen, our energetic data are in line with the other theoretical
values; moreover, the DFT/vdW-QHO-WF estimate of the interlayer 
distance turns out to be closer to the experimental reference value
(which, differently from the binding energy, is precisely determined)
than the distances predicted by the other theoretical schemes.

As our final application test we have considered the interaction 
of H$_2$ on Cu(111).
Adsorption processes on solid surfaces represent a very important topic
both from a fundamental point of view and to design and optimize 
countless material applications.
In particular, the adsorption of closed electron-shell particles, such as
rare-gas atoms and the H$_2$ molecule on metal
surfaces is prototypical\cite{Bruch} for ''physisorption'' processes,
characterized by an equilibrium between attractive, 
long-range van der Waals (vdW) interactions and short-range Pauli repulsion.
For the H$_2$ molecule on low-index Cu surfaces, accurate
physisorption data from experiment are available. 
Actually H$_2$ is the only molecule for which a detailed mapping of the
gas-surface interaction potential has been performed with resonance 
scattering measurements (see ref. \onlinecite{Lee2012} and references
therein). 

For the H$_2$-Cu(111) system we have modeled the metal surface
using a periodically-repeated hexagonal supercell,
with a $(\sqrt{3}\times \sqrt{3})R30^{\circ}$ structure and a surface slab
made of 15 Cu atoms distributed over 5 layers; moreover,  
we have adopted the same computational
approach of our previous study,\cite{PRB2013} but for the replacement of
the PW91\cite{PW91} functional with the PBE one, 
for the sake of uniformity with the other calculations performed 
with DFT/vdW-QHO-WF.
Similarly to the graphite case described above and as done in previous 
applications on adsorption
processes,\cite{silvsurf,silvmetodo,silvinter,ambrosetti,Ar-Pb}
we have also included the vdW interactions of the MLWFs of the 
physisorbed H$_2$ molecule not
only with the MLWFs of the underlying surface, within the reference supercell,
but also with a sufficient
number of periodically-repeated surface MLWFs (in any case, given the
$R^{-6}$ decay of the vdW interactions at large distances, the convergence with
the number of repeated images is rapidly achieved).
The binding energy has been evaluated for several 
adsorbate-substrate distances; then
the equilibrium distances and the corresponding binding energies 
have been obtained (as in refs. \onlinecite{Ar-Pb,PRB2013}) by fitting
the calculated points with the function: $A\,e^{-Bz}-C_3/(z-z_0)^3$.

In Table III DFT/vdW-QHO-WF results 
(DFT/vdW-WF-QHO$_{SL}$ and DFT/vdW-WF-QHO$_{TF}$ denote
the DFT/vdW-WF-QHO method with
metal screening effects included by the single-layer approximation
and the Thomas-Fermi scheme described above) are compared
to available theoretical and experimental estimates 
and to corresponding data obtained using our previous 
DFT/vdW-WF2s schemes\cite{PRB2013} (in that case using the PW91 reference
DFT functional), and other vdW-corrected DFT
approaches. 
As found in the previous studies\cite{Ar-Pb,PRB2013} 
the effect of the vdW-corrected schemes is 
a much stronger bonding than with a pure PBE scheme, 
with the formation of a clear minimum in the
binding energy curve at a shorter equilibrium distance.
Moreover, by comparing with unscreened data, obtained by
bare DFT/vdW-QHO-WF (we recall that
also the other vdW-corrected DFT methods do not take 
explicitly metallic screening
into account), we find that the effect of screening 
is substantial, leading to reduced binding energies and 
increased adparticle-substrate equilibrium distances.
It is reassuring that the results obtained by using the two different
recipes to describe metal screening are very similar.

Both the DFT/vdW-WF-QHO$_{SL}$ and DFT/vdW-WF-QHO$_{TF}$
binding energies turn out to slightly underestimate
the experimental value, although they are not worse than the predictions
of the vdW-DF, vdW-DF2 and rVV10 methods which instead tend to overestimate
it. Actually, the discrepancy with respect to the experiment is comparable with
the uncertainty associated to the approximate treatment of the 
metal-screening effect.\cite{PRB2013} 

The DFT/vdW-WF-QHO$_{SL}$ and DFT/vdW-WF-QHO$_{TF}$ equilibrium
distances of H$_2$ on Cu(111) are instead much better than with the other
methods, with the exception of rVV10. 
The same is true for the estimated C$_3$ coefficient if
comparison is done with the reference value of Vidali et al.\cite{Vidali}  

\section{Conclusions}
In summary, we have presented a scheme to
include the vdW interactions in DFT by combining the QHO model
with the MLWF technique.
The method has been applied to the S22 molecular database, and
also to extended systems, namely graphite and H$_2$ adsorbed on the
Cu(111) metal surface (in this case metal screening effects are taken into
account).
By comparing the results with those obtained by
other vdW-corrected DFT schemes the performances are satisfactory and 
turn out to be better than those of the previous DFT/vdW-WF 
and DFT/vdW-WF2 approaches, also based on the use of the MLWFs. 

\section{Acknowledgements}
We thank very much R. Sabatini for help in performing rVV10 calculations, and
A. Ambrosetti and A. Tkatechenko for useful discussions.

\vfill
\eject

\begin{table}
\caption{
Performance of different schemes
on the S22 database of intermolecular interactions.
The errors are measured with respect to basis-set
extrapolated CCSD(T) calculations of Takatani et al.\cite{Takatani}
Mean absolute relative errors (MARE in \%)
and mean absolute errors (MAE in
kcal/mol, and, in parenthesis, in meV) are reported.} 
\begin{center}
\begin{tabular}{|l|r|r|}
\hline
method & MARE & MAE \\ \tableline
\hline
DFT/vdW-WF-QHO &  7.7 & 0.71 [30.9] \\
DFT/vdW-WF     &  9.6 & 0.88 [38.2] \\
DFT/vdW-WF2    & 18.9 & 1.57 [68.1] \\
vdW-DF$^a$     & 17.0 & 1.22 [52.9] \\
vdW-DF2$^b$    & 14.7 & 0.94 [40.8] \\
VV10$^b$       &  4.4 & 0.31 [13.4] \\ 
rVV10$^c$      &  4.3 & 0.30 [13.0] \\ 
PBE+TS-vdW$^{d,e}$ & 10.3 & 0.32 [13.9] \\
PBE+MBD$^d$    &  6.2 & 0.26 [11.3] \\
PBE-D3$^{c,f}$ & 11.4 & 0.50 [21.7] \\ 
PBE$^{f,g}$    & 55.5 & 2.56[111.0] \\
\hline
\end{tabular}
\tablenotetext[1]{ref.\onlinecite{Cooper}.} 
\tablenotetext[2]{ref.\onlinecite{Vydrov}.} 
\tablenotetext[3]{ref.\onlinecite{Sabatini}.} 
\tablenotetext[4]{ref.\onlinecite{PNAS}.} 
\tablenotetext[5]{ref.\onlinecite{TS}.} 
\tablenotetext[6]{ref.\onlinecite{Grimme}.} 
\tablenotetext[7]{ref.\onlinecite{Zhao}.} 
\end{center}
\label{table1}
\end{table}
\vfill
\eject

\begin{table}
\caption{Interlayer binding energy, E$_b$, exfoliation energy, E$_e$, 
cleavage energy, E$_c$ (see text for the definitions), 
and interlayer distance, R, of graphite.}
\begin{center}
\begin{tabular}{|l|l|l|l|l|}
\hline
method                 &E$_b$ (meV)&E$_e$ (meV)&E$_c$ (meV)&R (\AA)\\ \tableline
\hline
DFT/vdW-WF-QHO         &-37        & -44        & -51        & 3.33 \\  
vdW-DF$^a$             &-45        & -48        & -50        & 3.60\\
``revised DFT''$^b$    &-34        &---         &---         & 3.50 \\ 
rVV10                  &-38        & -41        & -44        & 3.41 \\
TS-vdW+SCS$^c$         &-55        &---         &---         & 3.37 \\
QMC$^d$                &---        &---         & -60$\pm 5$ & 3.43$\pm 4$ \\
expt.$^e$              &---        &-35$\pm 10$ &---         & 3.34 \\
expt.$^f$              &-52$\pm 5$ &---         & -61$\pm 5$ & 3.34 \\
expt.$^g$              &-31$\pm 2$ &---         &---         & 3.34 \\
\hline
\end{tabular}
\tablenotetext[1]{ref.\onlinecite{Kack}.} 
\tablenotetext[2]{ref.\onlinecite{Rydberg}.} 
\tablenotetext[3]{ref.\onlinecite{Bucko}.}
\tablenotetext[4]{ref.\onlinecite{Spanu}.} 
\tablenotetext[5]{ref.\onlinecite{Benedict}.} 
\tablenotetext[6]{ref.\onlinecite{Zacharia}.} 
\tablenotetext[7]{ref.\onlinecite{Liu}.} 
\end{center}
\label{table2}
\end{table}
\vfill
\eject

\begin{table}
\caption{Binding energy E$_b$, see text for the definition,
equilibrium distance R, and estimated C$_3$ 
coefficient of H$_2$ on Cu(111). 
DFT/vdW-WF-QHO$_{SL}$ and DFT/vdW-WF-QHO$_{TF}$ denote 
the DFT/vdW-WF-QHO method with
metal screening effects included by the single-layer approximation
and the Thomas-Fermi scheme, respectively (see text).}
\begin{center}
\begin{tabular}{|l|r|c|r|}
\hline
method & E$_b$ (meV) &  R (\AA) & C$_3$ (meV\AA$^3$) \\ \tableline
\hline
DFT/vdW-WF-QHO &-58 & 3.03 & 1043 \\  
DFT/vdW-WF-QHO$_{SL}$&-23 & 3.47 &  647 \\  
DFT/vdW-WF-QHO$_{TF}$&-21 & 3.52 &  613 \\  
DFT/vdW-WF2s$^a$ & -36$\leftrightarrow$-26 & 3.40$\leftrightarrow$3.60 & 984$\leftrightarrow$1216 \\ 
vdW-DF$^{a,b}$     &-53 & 3.85 & 2310 \\
vdW-DF2$^{a,b}$    &-39 & 3.64 & 1097 \\
rVV10          &-41 & 3.52 & 1190 \\
DFT-D3$^{b,c}$     &-98 & 2.86 & --- \\
TS-vdW$^{b,d}$     &-66 & 3.20 & --- \\
PBE            & -6 & 4.10 & --- \\
ref.\onlinecite{Vidali}     & ---  & ---  & 673 \\
expt.$^e$      &-29$\pm 5$ & 3.52 & --- \\
\hline
\end{tabular}
\tablenotetext[1]{ref.\onlinecite{PRB2013}.}
\tablenotetext[2]{ref.\onlinecite{Lee2012}.}
\tablenotetext[3]{ref.\onlinecite{Grimme}.}
\tablenotetext[4]{ref.\onlinecite{TS}.}
\tablenotetext[5]{ref.\onlinecite{Lee2012}.}
\end{center}
\label{table3}
\end{table}
\vfill
\eject

\pagestyle{empty}

                      
\end{document}